\documentclass[prd, amsfonts, twocolumn, nofootinbib, showpacs]{revtex4}
\usepackage{graphicx, epsfig}
\usepackage{color}
\usepackage{amsmath}
\usepackage{amssymb}
\newcommand{\be}{\begin{equation}}
\newcommand{\ee}{\end{equation}}
\newcommand{\bea}{\begin{eqnarray}}
\newcommand{\eea}{\end{eqnarray}}

\newcommand{\gapp}{\mathrel{\raise.3ex\hbox{$>$}\mkern-14mu
\lower0.6ex\hbox{$\sim$}}}
\newcommand{\lapp}{\mathrel{\raise.3ex\hbox{$<$}\mkern-14mu
\lower0.6ex\hbox{$\sim$}}}
\def\bbox{{\,\lower0.9pt\vbox{\hrule \hbox{\vrule height 0.2 cm
\hskip 0.2 cm \vrule  height 0.2 cm}\hrule}\,}}

\begin{document}
\title{Maximal temperature in a simple thermodynamical system}
\author{De-Chang Dai$^1$, Dejan Stojkovic$^2$}
\affiliation{ $^1$Institute of Natural Sciences, Shanghai Key Lab for Particle Physics and Cosmology, and Center for Astrophysics and Astronomy, Department of Physics and Astronomy, Shanghai Jiao Tong University, Shanghai 200240, China}
\affiliation{ $^2$ HEPCOS, Department of Physics, SUNY at Buffalo, Buffalo, NY 14260-1500}

 %%%%%%%%%%%%%%%%%%%%%%%%%%%%%%%%%%%%%%%%%%%%%%%%%%%%%%%

\begin{abstract}
\widetext
Temperature in a simple thermodynamical system is not limited from above.
It is also widely believed that it does not make sense talking about temperatures higher than
the Planck temperature in the absence of the full theory of quantum gravity.
Here, we demonstrate that there exist a maximal achievable temperature in a system where particles obey the laws of quantum mechanics and classical gravity before we reach the realm of quantum gravity. Namely, if two particles with a given center of mass energy come at the distance shorter than the Schwarzschild diameter apart, according to classical gravity they will form a black hole. It is possible to calculate that a simple thermodynamical system will be dominated by black holes at a critical temperature which is about three times lower than the Planck temperature. That represents the maximal achievable temperature in a simple thermodynamical system.
\end{abstract}

%%%%%%%%%%%%%%%%%%%%%%%%%%%%%%%%%%%%%%%%%%%%%%%%%%

\maketitle
Newtonian mechanics does not impose any fundamental upper limit on velocities of  particles. While special relativity does limit velocities to the speed of light, there is still no fundamental upper limit on energies, which can be infinite. Since the temperature of a thermodynamical system is a measure of an average kinetic energy of particles in the system, special relativity does not impose an upper limit on the temperature either. Once we include gravity into consideration, situation becomes more complicated. In the absence of the full theory of quantum gravity, it is widely believed that the maximal temperature that make sense talking about is the Planck temperature \cite{sak}. At temperatures higher than that quantum gravity effects become very important and we simply do not know what happens in that regime.

In $1960$'s, Hagedorn studied the theory of hadron production and showed that at some finite temperature an increase of collision energy will increase the entropy of the system (i.e. number of states) rather than the temperature. The system will therefore be stuck at that temperature value \cite{Hagedorn:1965st,Frautschi:1971ij,Huang:1970iq}. However, this temperature does not represent an upper limit, it merely signals a phase transition where hadrons are effectively converted into a sea of quarks. But that quark matter can be further heated up.

In string theory, at high temperatures, the density of states grows exponentially indicating a phase transition at which very long strings are copiously produced \cite{Atick:1988si,Tye:1985jv,Alvarez:1985fw,Bowick:1985az}. By tuning the string tension one can make this transition happening at temperatures lower than the Planck temperature. This Hagedron temperature could represent a maximal achievable temperature because any increase in energy of the system would go into creating new stringy states rather than increasing the temperature (for an alternative point of view see \cite{Dienes:2005vw}).

In this letter we demonstrate that there is an upper limit on the temperature in a thermodynamical system before we enter the realm of quantum gravity, without using string theory of other quantum gravity approaches. The calculations are relatively straightforward and assumptions are justified.

We consider a thermodynamical system which consists of a certain number of particles in a box. The volume of the box is $V$. We assume that particles are in a thermal equilibrium with the temperature $T$. We define the fundamental constants which we will use throughout the paper:  $k_B$ is the Boltzmann constant, $c$ is the speed of light, $h$ is the Planck constant and $G$ is the Newton's gravitational constant. The quantum distribution function is either Bose-Einstein or Fermi-Dirac, or in the classical case Boltzmann distribution. We can write all three of them as
\begin{equation} \label{df}
f_i(E)=\frac{1}{\exp\left(\frac{E-\phi_i}{k_BT}\right) + p_i}
\end{equation}
The parameter $p_i$ takes the values $p_i =1$ or $p_i =-1$ if the $i$th particle is a fermion or a boson respectively, and $p_i=0$ if the $i$th particle satisfies Boltzmann distribution. $E$ is the particle energy, and $\phi_i$ is the $i$th particle's chemical potential. The total particle distribution is then
\begin{equation}
F(E)=\sum_i f_i(E) .
\end{equation}
As usual, the particle distribution function gives the probability that a particle occupies a certain state with a given energy and chemical potential at a given temperature.

To simplify the discussion, we consider that all the particles in the system are bosons of the same kind. At very high temperatures that we expect to find here, we can safely assume that the particles are massless, so $E=ck$, where $k$ is the magnitude of the momentum of the particle. This also sets the chemical potential $\phi = 0$.
We can calculate energy density $\rho_r$ for our case of a massless boson with one degree of freedom as
\begin{equation} \label{rho}
\rho_r =\int_0^{\infty}   F(E) ck \frac{d\vec{k}}{h^3} = \frac{4 k_B^4 \pi^5}{15 c^3 h^3} T^4,
\end{equation}
where $T$ is the temperature of radiation in the system, and $\frac{d\vec{k}}{h^3}$ is the element of the volume in the phase space.

Let's first review a crude argument based on global parameters of the system.
If the volume, $V$, occupied by particles is fixed, then the total mass of the system is $M= \rho_r V/c^2$. The Schwarzschild radius associated with this total mass is   $R_S =\frac{2GM}{c^2}$. The condition that the system is not within its own Schwarzschild radius is
\be \label{bhc}
2GM/c^2 \leq \left(\frac{3}{4 \pi} V\right)^{1/3} ,
\ee
where we assumed that the volume is spherically symmetric. From here we get the limit that the temperature must obey
\be \label{inf}
T \leq \frac{\left(3^{1/3} 5^{1/4} c^{7/4} h^{3/4}\right)}{\left( 2^{11/12} G^{1/4} k_B \pi^{4/3} V^{1/6}\right)} .
\ee
We can maximize this temperature by minimizing the volume of the system. Since the smallest volume that makes sense talking about is the Planck volume, $V_{\rm Pl} = (4\pi/3)(1.6 \times 10^{-35} m)^3$, this gives
\be
T \leq \frac{\left(3^{1/3} 5^{1/4} c^{7/4} h^{3/4}\right)}{\left( 2^{11/12} G^{1/4} k_B \pi^{4/3} V_{\rm Pl}^{1/6}\right)} = 1.1 \times 10^{32} K ,
\ee
 which is almost as high as the Planck temperature $T_{\rm Pl} = 1.42 \times 10^{32} K$. Interpretation of this result depends on how one looks at the physics at the Planck scale. It is true that Eq.~(\ref{inf}) allows infinite temperatures, if quantum gravity somehow allows for arbitrarily small volumes. But the corrections due to unknown physics become of the order one at the Planck scale, so one can't really extrapolate the Eq.~(\ref{inf}) beyond the Planck volume. Therefore, the maximal temperature will strongly depend on the scale where one makes the cut-off. In that sense, this argument is circular.

\begin{figure}[ht!]
   \centering
\includegraphics[width=10cm]{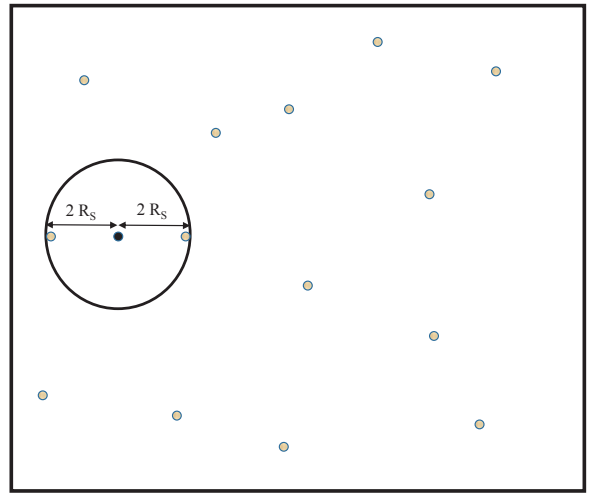}
\caption{If we single out one particle (black dot), a black hole will form if any other particle comes at the distance shorter than twice the Schwarzschild radius corresponding to their center of mass energy.}
\label{bh}
\end{figure}

The above calculations depend on global quantities of the system like its total volume and mass.
However, we can perform more sophisticated analysis that depends only on densities, and does not require an explicit input of global quantities.
Let's concentrate on local two-particle interactions.
According to the hoop conjecture \cite{hoop}, a black hole will form if two particles come to the distance which is shorter than twice the Schwarzschild radius corresponding to their center of mass energy (see Fig.~\ref{bh}). We can again define the Schwarzschild radius as
\begin{equation} \label{sr}
R_S =\frac{2GM}{c^2}
\end{equation}
but now $M$ is the center of mass energy of two particles \cite{Dai:2007ki} (as opposed to the total mass of the whole system like in the previous example) . Let's single out one particle with momentum $\vec{k}_1$.
The volume inside the Schwarzschild diameter of $2R_S$ is $V_S=(4\pi/3)  (2 R_S)^3$. The number of particles with momentum $\vec{k}_2$ inside this volume is
$V_S \times n(\vec{k}_2)$, where $n(\vec{k})$ is the number density of particles with momentum $\vec{k}$. The number density $n(\vec{k}_2)$ is obtained straight from the distribution
\be
dn(\vec{k}_2)= F(E_2)d^3\vec{k}_2 .
\ee
The element of the probability for a particle with momentum $\vec{k}_1$ to meet a particle with momentum $\vec{k}_2$ and create a black hole is then
%\begin{eqnarray}
%&&\frac{4\pi}{3}r_s^{3/2}F(E_2)\frac{d^3k_2}{h^3}\\
%&&M=\sqrt{m_1^2+m_2^2+2E_1E_2/c^4-2k_1 k_2 /c^2 \cos\theta}\\
%&& E_2 =\sqrt{k_2^2c^2 +m_i^2c^4}\\
%&& E_1 =\sqrt{k_1^2c^2 +m_i^2c^4}
%\end{eqnarray}
\be \label{pbh}
dP_{\rm BH} = \frac{4\pi}{3}(2R_S)^{3}F(E_2)\frac{d^3\vec{k}_2}{h^3}
\ee
The center of mass energy of these two particles with momenta $\vec{k}_1$ and $\vec{k}_2$ is
\be \label{come}
M=\sqrt{m_1^2+m_2^2+2E_1E_2/c^4-2k_1 k_2 /c^2 \cos\theta}
\ee
where $\theta$ is the angle between $\vec{k}_1$ and $\vec{k}_2$. $E_1$ and $E_2$ are the energies of particles with $\vec{k}_1$ and $\vec{k}_2$ respectively
\begin{eqnarray}
 &&E_2 =\sqrt{k_2^2c^2 +m_1^2c^4}\\
&& E_1 =\sqrt{k_1^2c^2 +m_2^2c^4}
\end{eqnarray}
where $m_1$ and $m_2$ are masses of particles with $\vec{k}_1$ and $\vec{k}_2$.

We can calculate the number density of black holes from the probability that two particles make a black hole in collision (given in Eq.~(\ref{pbh})), and the particle distribution functions
\begin{equation} \label{nbh}
n_{\rm BH}=\frac{1}{2}\int  \frac{4\pi}{3}(2R_S)^{3}F(E_2)F(E_1)\frac{d^3\vec{k}_1}{h^3} \frac{d^3\vec{k}_2}{h^3}
\end{equation}
We added $2$ in the denominator, because $\vec{k}_1$ and $\vec{k}_2$ are exchangeable. The element of the volume in momentum space is
\be
d^3\vec{k}_{1,2} = k_{1,2}^2 \sin \theta_{1,2} d\theta_{1,2} d\varphi_{1,2} dk_{1,2}.
 \ee
 Integration over $\varphi_{1,2}$ is trivial, but integration over $\theta_{1,2}$ requires more attention. The integrand in Eq.~(\ref{nbh}) depends on $\theta$ which is the angle between the colliding particles $\vec{k}_{1}$ and $\vec{k}_{2}$. We can always rotate a coordinate system in which we measure the angle $\theta$ to align with the new $z$-axis, for which we chose the direction of $\vec{k}_{1}$. Then, the angle $\theta$ becomes $\theta_2$. Integration over $\theta_1$ and $\varphi_1$ will give the usual $4\pi$.
 %We therefore have
%\be
%n_{\rm BH} = \frac{1}{2}\frac{4}{3}\frac{\pi}{h^6} \int_{k_1=0}^\infty \int_{k_2=0}^\infty \int_{\theta_2=0}^\pi
 %(2R_S)^3F(E_2)F(E_1)4\pi 2\pi k_1^2 k_2^2 \sin \theta_2 dk_1 dk_2 d\theta_2  .
%\ee
Assuming again massless particles, after substituting Eqs.~(\ref{df}), (\ref{sr}) and (\ref{come}), we get
\begin{eqnarray} \label{nbhf}
&& n_{\rm BH} =  \frac{2048 \sqrt{2} \pi^3 }{3 h^6 c^9}  \times \\
&& \times \int_{k_1=0}^\infty \int_{k_2=0}^\infty \int_{\theta_2=0}^\pi
\frac{ k_1^{7/2} k_2^{7/2} \left(1 - \cos \theta_2 \right)^{3/2} \sin \theta_2 dk_1 dk_2 d\theta_2 }{ \left(e^{\frac{c k_1}{k_B T}} -1 \right) \left(e^{\frac{c k_2}{k_B T}} -1 \right)}  \nonumber
\end{eqnarray}

Integration in Eq.~(\ref{nbhf}) can be performed exactly to give
\be \label{nbhfn}
n_{\rm BH} = \frac{94080 G^3 k_B^9 \pi^4 \zeta[9/2]^2}{c^{18} h^6}T^9
\ee
where $\zeta$ represents the Riemann zeta function.

We need to compare the number density of black holes with the number density of particles defined as
 \begin{equation} \label{n}
n =\int_0^{\infty}   F(E) \frac{d\vec{k}}{h^3} .
\end{equation}
Again, integration in Eq.~(\ref{n}) can be performed exactly to give
\be
n = \frac{8 k_B^3 \pi \zeta[3]^2}{c^{3} h^3} T^3
\ee
We see that the number density of black holes grows with temperature much faster than the number density of particles ($T^9$ vs. $T^3$), but there is a huge suppression factor coming from the weakness of gravity that black holes have to overcome.
At the critical temperature, $T_C$, the number density of black holes will become greater than the number density of particles, i.e. $n_{\rm BH} \geq n$.
This yields the critical temperature
\be \label{tc1}
T_C =   \frac{c^{5/2} \sqrt{h} \left(\zeta[3]/15\right)^{1/6}}{2^{2/3} \sqrt{\pi G} k_B \left(7 \zeta[9/2]\right)^{1/3}} = 4.27 \times 10^{31} K .
\ee
For comparison, the Planck temperature is $T_{\rm Pl} = 1.42 \times 10^{32} K$. In fact, if we use the definition of the Planck temperature $T_{\rm Pl} = \sqrt{(hc^5/(2 \pi G k_B^2)}$, our result is
\be \label{tc2}
T_C =   \frac{\left(\zeta[3]/30\right)^{1/6}}{\left(7 \zeta[9/2]\right)^{1/3}} \times T_{\rm Pl} = 0.30 \times T_{\rm Pl} .
\ee

\begin{figure}[h!]
   \centering
\includegraphics[width=9cm]{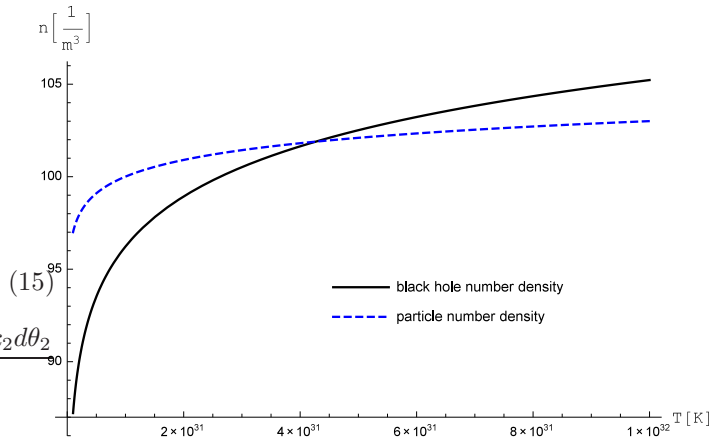}
\caption{ Black hole number density $n_{\rm BH}$ (solid line) and particle number density $n$ (dashed line ) as a function of temperature in a thermodynamical system. Densities are plotted in units of ${\rm meter}^{-3}$. The black hole number density, $n_{\rm BH}$, becomes dominant at the temperature $T\approx 4.27 \times 10^{31} K$, which is lower than the Planck temperature $T_{\rm Pl} = 1.42 \times 10^{32} K$. This temperature represents the maximal achievable temperature in a simple thermodynamic system.}
\label{tem}
\end{figure}

To show more details, we plot the black hole and particle number density as a function of temperature in Fig.~\ref{tem}. We see that at $T < 4.27 \times 10^{31} K$ particles dominate the system, while at  $T > 4.27 \times 10^{31} K$ black holes dominate the system. This defines the maximum temperature that can be achieved in the system above which particles cannot effectively exist anymore.

There is one more thing left to check. Once formed, a black hole will start evaporating. A small black hole might disappear quickly and get converted back into radiation. So we have to check that this does not happen above $T_C$. The black hole temperature is
\begin{equation} \label{tbh}
T_{\rm BH}=\frac{h c^3}{16 \pi^2 G M k_B} ,
\end{equation}
where $M$ is the black hole mass (in our case the center of mass energy of two colliding particles). The rate of mass loss due to Hawking evaporation is
\be \label{er}
\frac{d(Mc^2)}{dt} =  A_{\rm BH} \frac{ \pi^5 k_B^4}{15h^3 c^2} T_{\rm BH}^4 ,
\ee
where $A_{\rm BH} = 4 \pi R_S^2 $ is the area of the black hole, while the constant $a = 4 \sigma /c$ where  $\sigma = \frac{2 \pi^5 k_B^4}{15h^3 c^2}$ is the Stefan-Boltzmann constant.
Simultaneously with evaporation, the black hole will start accreting material from the environment. At high temperatures most of matter will be in form of radiation. The rate of accretion will therefore be
\be
\frac{d(Mc^2)}{dt} = \frac{c}{4} A_{\rm BH} \rho_r ,
\ee
where $\rho_r$ is the energy density of radiation, and we assumed that the efficiency of accretion is one.
The energy density $\rho_r$ was calculated in Eq.~(\ref{rho}) as
\begin{equation}
\rho_r = \frac{4 k_B^4 \pi^5}{15 c^3 h^3} T^4 ,
\end{equation}
where $T$ is the temperature of radiation in the system.
This gives the accretion rate
\be \label{ar}
\frac{dM}{dt} = A_{\rm BH} \frac{4 k_B^4 \pi^5}{15 c^4 h^3} T^4 .
\ee
 We define the equilibrium temperature of the system, $T_{\rm eq}$, when the rate or evaporation in Eq.~(\ref{er}) and the rate or accretion in Eq.~(\ref{ar}) are equal. Therefore
\be
T_{\rm eq} =  T_{\rm BH} ,
\ee
taking into account that $\sigma$ in Eq.~(\ref{er}) was calculated for photons which have two degrees of freedom.   In fact, this result is guaranteed by the zeroth law of thermodynamics.

We want to ensure that the black holes in the system above $T_C$ have large enough mass so that their temperature is below $T_C$ (otherwise their evaporation will dominate accretion). So we impose condition
\be \label{con}
T_{\rm BH} < T_C.
\ee
The black hole temperature $T_{\rm BH}$ depends on the collision angle $\theta$ through the center of mass energy as $M^2= 2k_1 k_2 (1-\cos\theta )/c^2$.
As $\theta$ goes from  zero to $\pi$, the center of mass energy goes from zero to $M_{\rm max}^2= 4 k_1 k_2 / c^2$. Obviously, we want to exclude very light black holes which have large temperatures. Therefore, the condition in Eq.~(\ref{con}) will impose the lower limit on $\theta$
\be
\theta_{\rm min} =  \cos^{-1} \left[1- \frac{c^8 h^2}{512 G^2 k_1 k_2 k_B^2 \pi^4 T_C^2} \right]
\ee
We can now perform numerical integration in Eq.~(\ref{nbhf}) where the angle $\theta$ goes over $(\theta_{\rm min}, \pi)$. One can check that the plot is practically identical as one in Fig.~\ref{tem}. This indicates that most of the black holes at $T>T_C$ have temperatures lower than $T_C$, and excluding those with higher temperatures did not change the result at all.

Here we showed the results only for the Bose-Einstein distribution function. However, we also explicitly checked that the results do not visibly change for the Fermi-Dirac and Boltzmann distributions.

According to what we found so far, the following sequence of events happens as the temperature in the system increases. We assume that the system at the critical temperature is not already within its own Schwarzschild radius. When the temperature of the system is below critical, $T < T_C$, the number of black holes is small and particles dominate the system. However, when the temperature of the system is above critical, $T > T_C$, we have more black holes than particles. In addition, their temperature is lower than that of the system, so accretion wins and black holes grow in size, which in turn destroys the initial system of particles. When the whole system turns into a few or perhaps just one large black hole, a turning point is reached. Then evaporation will start wining until the critical temperature is achieved again, and the whole cycle might be repeated many times. Thus, for every thermodynamical system, there should be a maximal temperature that can not be surpassed.

It is important to note that the critical temperature, $T_C$, we found here is three times lower than the Planck temperature. This means that our calculations are under control and quantum gravity corrections will not ruin the result. We indeed used the process of black hole production by two particles, but this is purely classical process as long as the energy densities and temperatures are below the Planck scale.

If the system has a constant volume $V$, then one can find from the condition in Eq.~(\ref{bhc}) that $V$ must be at most $10^2$ larger than the Planck volume in order not to be within its own Schwarzschild radius at $T = 4.27 \times 10^{31} K$. However, in a dynamical situation where the volume is not constant, the condition in Eq.~(\ref{bhc}) does not hold and the volume might be much larger. In fact, in the context of expanding cosmology, the volume $V$ might be as large as the whole Hubble volume.

One could generalize our calculations in a number of ways. One can for example consider different types of particles in the mixture, and also include interactions between them.
One might also perform calculations in the context of a concrete cosmological model.

To conclude, using only quantum mechanics, classical gravity and statistical physics we demonstrated that the maximal achievable temperature in a simple thermodynamic system is about three times lower than the Planck temperature. It is important to note that we did not assume any a priori cutoff in our calculations.  We started with fundamental constants $G$, $h$, $k$, and $c$, and along the way the Planck temperature naturally factored out in eqs. (\ref{tc1}) and (\ref{tc1}). Our calculations are performed in regime where eventual corrections due to unknown physics are relatively small.

%\section{Comparison with previous work}

At the end, we comment on a previous related work.
In \cite{gpy}, the authors considered the black hole nucleation rate in flat space-time. They showed that the Jeans instability arises as a tachyon in the graviton propagator when small perturbations about hot flat space are considered. The vacuum decay rate due to black hole nucleation is given in their equation 5.40. This is in essence an instanton effect which is quantum in its nature, and is exponentially suppressed below the Planck scale. This is an interesting result showing that even a hot flat space-time is ultimately unstable to nucleating black holes. In contrast, our mechanism for black hole production is different. It is a collision of two particles whose center of mass energy is concentrated in a region smaller than the corresponding Schwarzschild radius. This process is classical in its nature and it is not described by instantons.

Another paper \cite{pw} deals with the similar question of black hole production in a thermal system.   The authors considered a system with a fixed volume and energy, and estimated the probability that a thermal fluctuation will lead to overdensities which can produce black holes. The rate of black hole production is given in their Eq.~(12). To calculate the maximal temperature in the system it is not important how quickly the black holes are formed, all that is needed is that they do not evaporate once they are formed and that at certain point dominate the system.  Thus, one cannot directly compare their Eq.~(12) with our Eq.~(\ref{nbhf}) and Eq.~(\ref{nbhfn}). One could say that their approach is top-down (or global), while ours is bottom-up (or local). More precisely, we calculate probabilities to form black holes in different ways.  We calculate the probability from the particle energy distribution function. Particles are treated individually.  We concentrate on the two-body collisions since they give the dominant contribution.  In contrast, the authors of \cite{pw} calculate the probabilities from thermal dynamics. Their probability comes from entropy of the system (Eq.~(4) in their paper).  Their effect is average, while our effect is local.

However, the above mentioned papers papers are not making any connection between the black hole nucleation and the existence of a maximal temperature in a thermodynamically system.

Here, we briefly extend the calculations from \cite{pw} in order to get an analog expression for the maximal temperature in the system. According to \cite{pw}, the black hole creation rate per unit volume in an energy band between $E$ and $E+dE$, for a low temperature gas is

\begin{equation}
R(E)dE \approx \frac{1}{k_B T} \exp(-\frac{E}{k_B T})l_p^{-3}t_p^{-1}
\end{equation}
where $l_p\approx 10^{-33}\text{cm}$ and $t_p\approx 10^{-43}\text{s}$ (this is Eq.~13 in \cite{pw}). Within the approximations that the authors use, most of the black holes are created with mass around Planck mass for temperatures $T\ll T_p$. The time scale for creation of such black holes within some volume $V$ is
\begin{equation}
t\sim \frac{l_p^3}{V}\exp(\frac{M_p}{k_B T})t_p ,
\end{equation}
where $M_p$ is the Planck mass. A characteristic black hole evaporation time measured in seconds is
\begin{equation}
t_{ev}= \frac{5120\pi G^2 M^3}{\hbar c^4}=8.67\times 10^{-40}\frac{M^3}{M_p^3} .
\end{equation}
For a black hole to survive in some volume $V$, the characteristic creation time scale must be shorter than the characteristic evaporation time scale, i.e. $t<t_{ev}$. From this condition we get
\begin{equation}
T>\frac{M_p}{k_B}\frac{1}{9.06+\ln(V/l_p^3) }
\end{equation}
Here $t_{ev}$ is calculated for $M=M_p$. Above this temperature, a created black hole will most like survive, so one might roughly interpret it as the maximal temperature in the system.

%Here $t_{ev}$ is estimated under $M=M_p$ that is an underestimation because the black holes are heavier than Planck mass, $M>M_p$ . A thermal black hole can be created statically %below the Planck temperature. If the volume is big enough, the temperature can be even lower.

While this result is numerically close to ours, its interpretation is somewhat different. As we mentioned, the approach in \cite{pw} utilizes average properties of the system like its entropy. Entropy alone cannot give the particle distribution function. The black hole creation rate gives the (inverse) characteristic time for creation of a single black hole in some volume. The black hole in question may evaporate, leave the volume, and be replaced by a new black hole. This is different from our approach which directly uses the particle distribution function. We found the critical temperature at which there are more black holes than particles. The rate at which the black holes are produced is not crucial for this.

\begin{acknowledgments}
D.C Dai was supported by the National Science Foundation of China (Grant No. 11433001 and 11447601), National Basic Research Program of China (973 Program 2015CB857001), No.14ZR1423200 from the Office of Science and Technology in Shanghai Municipal Government and the key laboratory grant from the Office of Science and Technology in Shanghai Municipal Government (No. 11DZ2260700) and  the Program of Shanghai Academic/Technology Research Leader under Grant No. 16XD1401600. D.S. was partially supported by the US National Science Foundation, under Grant No. PHY-1417317.
\end{acknowledgments}


\begin{thebibliography}{99}

\bibitem{sak}
A.D. Sakharov, JETP Lett. 3 (1966) 288

%\cite{Hagedorn:1965st}
\bibitem{Hagedorn:1965st}
  R.~Hagedorn,
  %``Statistical thermodynamics of strong interactions at high-energies,''
  Nuovo Cim.\ Suppl.\  {\bf 3}, 147 (1965).
  %%CITATION = NUCUA,3,147;%%
  %1256 citations counted in INSPIRE as of 06 Dec 2015
%\cite{Frautschi:1971ij}
\bibitem{Frautschi:1971ij}
  S.~C.~Frautschi,
  %``Statistical bootstrap model of hadrons,''
  Phys.\ Rev.\ D {\bf 3}, 2821 (1971).
  doi:10.1103/PhysRevD.3.2821
  %%CITATION = doi:10.1103/PhysRevD.3.2821;%%
%\cite{Huang:1970iq}
\bibitem{Huang:1970iq}
  K.~Huang and S.~Weinberg,
  %``Ultimate temperature and the early universe,''
  Phys.\ Rev.\ Lett.\  {\bf 25}, 895 (1970).
  doi:10.1103/PhysRevLett.25.895
  %%CITATION = doi:10.1103/PhysRevLett.25.895;%%


\bibitem{Atick:1988si}
  J.~J.~Atick and E.~Witten,
  %``The Hagedorn Transition and the Number of Degrees of Freedom of String Theory,''
  Nucl.\ Phys.\ B {\bf 310}, 291 (1988).
  doi:10.1016/0550-3213(88)90151-4
  %%CITATION = doi:10.1016/0550-3213(88)90151-4;%%

%\cite{Tye:1985jv}
\bibitem{Tye:1985jv}
  S.~H.~H.~Tye,
  %``The Limiting Temperature Universe and Superstring,''
  Phys.\ Lett.\ B {\bf 158}, 388 (1985).
  doi:10.1016/0370-2693(85)90438-1
  %%CITATION = doi:10.1016/0370-2693(85)90438-1;%%

 %\cite{Alvarez:1985fw}
\bibitem{Alvarez:1985fw}
  E.~Alvarez,
  %``Strings At Finite Temperature,''
  Nucl.\ Phys.\ B {\bf 269}, 596 (1986).
  doi:10.1016/0550-3213(86)90514-6
  %%CITATION = doi:10.1016/0550-3213(86)90514-6;%%

  %\cite{Bowick:1985az}
\bibitem{Bowick:1985az}
  M.~J.~Bowick and L.~C.~R.~Wijewardhana,
  %``Superstrings at High Temperature,''
  Phys.\ Rev.\ Lett.\  {\bf 54}, 2485 (1985).
  doi:10.1103/PhysRevLett.54.2485
  %%CITATION = doi:10.1103/PhysRevLett.54.2485;%%
  %172 citations counted in INSPIRE as of 06 Dec 2015

%\cite{Dienes:2005vw}
\bibitem{Dienes:2005vw}
  K.~R.~Dienes and M.~Lennek,
  %``Re-identifying the Hagedorn transition,''
  hep-th/0505233.
  %%CITATION = HEP-TH/0505233;%%


\bibitem{hoop}
K. Thorne,  ``{\it Black Holes and Time Warps: Einstein's Outrageous Legacy}", W. W. Norton $\&$ Company; Reprint edition, January 1, 1995. ISBN 0-393-31276-3.

%\cite{Dai:2007ki}
\bibitem{Dai:2007ki}
  D.~C.~Dai, G.~Starkman, D.~Stojkovic, C.~Issever, E.~Rizvi and J.~Tseng,
  %``BlackMax: A black-hole event generator with rotation, recoil, split branes, and brane tension,''
  Phys.\ Rev.\ D {\bf 77}, 076007 (2008)
  doi:10.1103/PhysRevD.77.076007
  [arXiv:0711.3012 [hep-ph]].
  %%CITATION = doi:10.1103/PhysRevD.77.076007;%%
  %125 citations counted in INSPIRE as of 29 Nov 2015


\bibitem{gpy}  D. J. Gross, M. J. Perry, and L. G. Yaffe,
Phys. Rev. D {\bf 25}, 330 (1982)

\bibitem{pw}  T. Piran and R. M. Wald,
Phys. Lett. A {\bf 90}, 20  (1982)

\end{thebibliography}
\end{document}